\documentclass[sn-mathphys-ay]{sn-jnl}

\usepackage{graphicx}%
\usepackage{multirow}%
\usepackage{amsmath,amssymb,amsfonts}%
\usepackage{amsthm}%
\usepackage{mathrsfs}%
\usepackage[title]{appendix}%
\usepackage{xcolor}%
\usepackage{textcomp}%
\usepackage{manyfoot}%
\usepackage{booktabs}%
\usepackage{algorithm}%
\usepackage{algorithmicx}%
\usepackage{algpseudocode}%
\usepackage{listings}%

\usepackage{enumitem}%
\usepackage{braket}%
\usepackage{subcaption}%
\usepackage{comment}%
\usepackage{bm}%

\theoremstyle{thmstyleone}%
\theoremstyle{thmstyletwo}%

\theoremstyle{thmstylethree}%

\begin{document}

\title[QAE classification]{Quantum autoencoders for image classification}

\author*[1]{\fnm{Hinako} \sur{Asaoka}}\email{asaoka.hinako@is.ocha.ac.jp}

\author[1,2]{\fnm{Kazue} \sur{Kudo}}
\equalcont{These authors contributed equally to this work.}

\affil*[1]{\orgdiv{Department of Computer Science}, \orgname{Ochanomizu University}, \orgaddress{\street{2-1-1 Otsuka}, \city{Bunkyo-ku}, \postcode{112-8610}, \state{Tokyo}, \country{Japan}}}

\affil[2]{\orgdiv{Graduate School of Information Sciences}, \orgname{Tohoku University}, \orgaddress{\street{6-3-09 Aoba, Aramaki-aza Aoba-ku}, \city{Sendai}, \postcode{980-8579}, \state{Miyagi}, \country{Japan}}}

\abstract{Classical machine learning often struggles with complex, high-dimensional data. Quantum machine learning offers a potential solution, promising more efficient processing. The quantum convolutional neural network (QCNN), a hybrid algorithm, fits current noisy intermediate-scale quantum hardware. However, its training depends largely on classical computation. Future gate-based quantum computers may realize full quantum advantages. In contrast to QCNNs, quantum autoencoders (QAEs) leverage classical optimization solely for parameter tuning. Data compression and reconstruction are handled entirely within quantum circuits, enabling purely quantum-based feature extraction. This study introduces a novel image-classification approach using QAEs, achieving classification without requiring additional qubits compared with conventional QAE implementations. The quantum circuit structure significantly impacts classification accuracy. Unlike hybrid methods such as QCNN, QAE-based classification emphasizes quantum computation. Our experiments demonstrate high accuracy in a four-class classification task, evaluating various quantum-gate configurations to understand the impact of different parameterized quantum circuit (ansatz) structures on classification performance. An ideal situation in which quantum circuits yield perfect results is assumed, and a numerical simulator of quantum circuit measurements is used. Specifically, noise-free conditions are considered, and simulations are performed using a statevector simulator to model the quantum system with full amplitude precision. Our results reveal that specific ansatz structures achieve superior accuracy. Moreover, the proposed approach achieves performance comparable to that of conventional machine-learning methods while significantly reducing the number of parameters requiring optimization. These findings indicate that QAEs can serve as efficient classification models with fewer parameters and highlight the potential of utilizing quantum circuits for complete end-to-end learning.}




\maketitle

\section{Introduction}\label{sec1}

Quantum machine learning (QML) is an emerging field exploring the application of quantum computers to machine-learning tasks. QML leverages the intrinsic parallelism of quantum computation, offering the potential for more efficient processing of complex data compared to classical machine-learning methods \citep{Biamonte_2017, Tycola_2023}. Shor's algorithm and Grover's algorithm have demonstrated that quantum computers can efficiently solve problems that require impractical computation time using classical computers \citep{Shor_1997, Zhang_2020}. In more advanced applications, approaches have been proposed to replace machine learning algorithms—such as autoencoders, neural networks, reinforcement learning, adversarial learning, and transformers, whose effectiveness has already been confirmed in classical computing experiments—with algorithms to be executed on quantum computers \citep{Romero_2017, Cong_2019, Henderson_2019, Lu_2020, He_Liang_2021, Song_2024, Cherrat_2024}. Several QML studies have shown that the optimization of information from complex feature spaces may improve learning accuracy and minimize the number of parameters compared to classical methods \citep{Adhikary_2020, Chakraborty_2020, Wei_2022}. Although current noisy intermediate-scale quantum (NISQ) devices are limited by qubit count and high noise, various hybrid quantum-classical approaches, such as quantum circuit learning, have been proposed to enable practical QML applications \citep{Mitarai_2018}. A prominent example is the quantum convolutional neural network (QCNN), which integrates gate-based quantum computing with conventional machine-learning models \citep{Cong_2019, Henderson_2019}. Classical convolutional neural networks (CNN), widely used in image classification for their ability to extract abstract features through convolution, achieve robust accuracy by learning generalized image representations \citep{Khan_2020}. QCNN aims to enhance processing speed and accuracy by replacing classical convolution operations with quantum circuits. However, because only a portion of the convolutional process occurs on a quantum circuit, with the remainder handled by a classical computer, QCNN remains a hybrid algorithm well-suited to the limitations of NISQ-era hardware. This hybrid approach attempts to capitalize on the benefits of quantum computing while relying on classical processing for robustness \citep{Wei_2022, Matic_2022, Hur_2022, Hassan_2024, Senokosov_2024}. Critically, the dependence on classical algorithms remains high, as only a fraction of the learning process is performed quantum mechanically.

While hybrid methods such as QCNN are pragmatic for current technology, developing approaches where most of the learning process occurs within quantum circuits is also essential. Although such algorithms may not be immediately practical in the NISQ era, future advancements in large-scale, gate-based quantum computing promise to unlock the full potential of quantum effects such as superposition and entanglement in machine learning. This could lead to significantly faster learning and greater model expressiveness than classical approaches. A promising avenue for quantum feature learning is the quantum autoencoder (QAE) \citep{Romero_2017}. QAEs are generative models that compress input data into fewer qubits and then reconstruct the original information by reversing the trained quantum circuit. While classical optimization is used for parameter tuning in QAE training, data compression and reconstruction are performed entirely within the quantum circuit. This demonstrates the capability of quantum circuits alone to effectively extract essential features. Prior research has primarily employed QAEs for data compression and reconstruction, including applications in data preprocessing, feature transformation, and anomaly detection, all leveraging their compression-reconstruction capabilities \citep{Srikumar_2021, Mangini_2022, Ngair_2022, Sakhnenko_2022, Zhu_2023, Wang_2024}. However, to the best of our knowledge, no prior work has demonstrated the application of QAEs to other tasks.

Therefore, this study proposes a novel approach applying QAEs to image classification, demonstrating that QAEs, as generative models, can perform tasks beyond data compression and reconstruction. In a previous study, we successfully applied nonnegative/binary matrix factorization (NBMF) to image classification, leveraging binary optimization for feature compression \citep{O_Malley_2018, Asaoka_2023}. This resulted in faster convergence and higher accuracy compared to other machine-learning methods. Similarly, we anticipate that QAEs will offer several advantages in image classification. In our proposed QAE-based image classification, the label information is incorporated during training. Traditional QAE training optimizes quantum circuit parameters to minimize reconstruction error between input images and their reconstructed outputs. Our approach encodes both image and label information into the quantum circuit and optimizes parameters to minimize the error between the reconstructed output and label information. This allows the trained circuit to predict class labels for unseen test images. Importantly, this method enables image classification without increasing the number of qubits compared to conventional QAE implementations. While parameter optimization uses classical algorithms, the quantum circuit transforms the quantum representation of images and labels. Consequently, the quantum circuit structure significantly influences classification accuracy. Unlike hybrid models such as QCNN, which perform only a portion of the computation quantum mechanically, QAE-based classification emphasizes quantum computation in machine learning.

This study experimented with multiple qubit control configurations to investigate the effect of different ansatz structures within the parameterized quantum circuit of the QAE. We demonstrate that specific ansatz structures achieve high classification accuracy and analyze the underlying reasons for their effectiveness. Additionally, we compare the performance of QAE classification with that of other classical machine-learning models and QCNN under identical data conditions. We identify scenarios where QAEs outperform classical approaches and those wherein their learning efficiency is lower, providing insights into the potential advantages of QML.
The key contributions of this paper are as follows:
\begin{enumerate}
\item It proposes the application of QAE to image-classification tasks.
\item It identifies effective ansatz structures for QAE-based image classification.
\item It demonstrates that QAE-based classification achieves comparable accuracy to classical methods with significantly fewer parameters.
\end{enumerate}

\section{Related work}\label{sec2}

\subsection{Autoencoders}
An autoencoder is a classical generative model designed for dimensionality reduction and reconstruction of input data \citep{Hinton_2006}. It achieves this through an unsupervised neural network. Because the primary goal is input data reconstruction, the output maintains the same dimensionality as the input. This study considers a three-layer autoencoder structure: an input encoder, a latent layer, and an output decoder. The centrally located latent layer has fewer dimensions than the input layer, facilitating data compression. Network parameters are optimized during training to minimize the discrepancy between input and output data, ensuring accurate reconstruction. This training process enables the extraction of essential features within the latent layer, which are crucial for data reconstruction.

Autoencoders are used in various applications. In anomaly detection, they are trained solely on normal data and detect outliers based on elevated reconstruction errors for abnormal inputs \citep{Gong_2019, Chen_2018}. Autoencoder-based anomaly detection has been applied in industrial domains for identifying abnormal conditions, such as system failures or equipment malfunctions \citep{Jianbo_2024, Harrou_2024}. Autoencoders have also been utilized for data processing techniques, such as augmentation, reconstruction, and noise reduction, to improve the performance of machine learning \citep{Zhiqiang_2017, Majumdar_2019, Jiang_2022, Torabi_2023, Akkem_2024}. As a machine learning method, autoencoders are used in unsupervised learning to extract features from data. Image classification tasks are solved by training a classifier with the extracted features and class information \citep{Gogoi_2017, Luo_Wei_2018, Zhou_2019, Sun_2019}.

\subsection{Quantum circuit learning (QCL)}\label{sec:QCL}
QCL is a hybrid classical-quantum machine-learning framework \citep{Mitarai_2018}. It involves encoding input data into a quantum circuit and employing classical optimization algorithms to iteratively adjust circuit parameters to achieve the desired output. QCL aims to leverage the exponentially large state space of quantum systems for machine learning, potentially solving tasks that are intractable for classical algorithms and demonstrating quantum advantage. Training data with QCL begins by encoding the data into a quantum state, $\ket{\psi_{\text{in}}}$, through a unitary transformation. Applying a parameterized quantum circuit ansatz, $U(\bm{\theta})$, to the input state yields the output state $U(\bm{\theta})\ket{\psi_{\text{in}}}$. In supervised machine-learning scenarios, a cost function is defined in terms of the measurement results of the output state and the training data. The desired circuit is trained by iteratively optimizing the parameter $\bm{\theta}$ to minimize this cost function.

Quantum convolutional neural networks (QCNNs) are learning methods utilizing quantum circuit learning (QCL), in which parts of classical CNNs are replaced with parameterized quantum circuits. Initial approaches have been proposed to compute the convolution and pooling operations in classical CNNs using quantum computation \citep{Cong_2019, Hur_2022}. However, under the limitations in the scale and stability of quantum computation, many widely adopted methods aim to replace only part of the CNN with quantum computation. Introducing quantum circuits into a portion of the convolutional layers enables extreme reduction of quantum computation, making it suitable for the NISQ era \citep{Henderson_2019, Matic_2022}. There are also approaches in which features extracted by quantum convolution are input into classical fully connected networks \citep{Wei_2022, Shui_Yuan_2023, Senokosov_2024}. QCNNs have been used to solve image classification tasks on the MNIST dataset, achieving high classification accuracy \citep{Hur_2022, Shui_Yuan_2023, Senokosov_2024}. Furthermore, as more advanced and practical tasks, QCNNs have been applied to medical image analysis, where they have demonstrated high accuracy and reduced computational costs \citep{Matic_2022, Houssein_2022, Hassan_2024, Cherrat_2024}. In addition, QCNNs have shown high accuracy in applications such as financial forecasting and attack risk detection, indicating the potential to contribute to solving business challenges \citep{Egger_2020, Thakkar_2024, Song_2024, Xiong_2025}. However, it should be noted that these methods achieve high accuracy by increasing dependence on classical machine learning. In approaches that replace part of the CNN with quantum computation, the majority of the learning process is performed by classical machine learning methods. Even in methods where convolution and pooling are fully handled by quantum circuits, classical dimensionality reduction techniques such as PCA or autoencoders are often applied to compress the classical data before feeding it into the quantum circuits \citep{Hur_2022}. It is important to recognize that effects other than quantum circuit learning also contribute to achieving high accuracy.

\subsection{Quantum autoencoders}
\begin{figure}[t]
\centering
\includegraphics[width=0.5\textwidth]{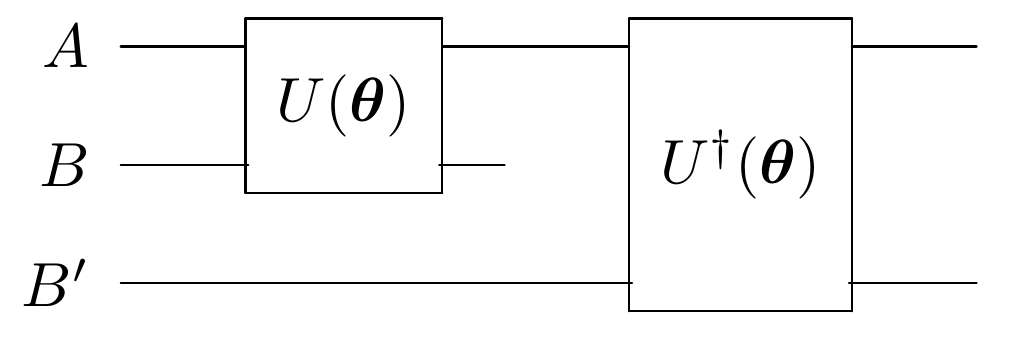}
\caption{Structure of a conventional quantum autoencoder}\label{fig:QAE_before_swap}
\end{figure}

A QAE is an algorithm that employs QCL to construct an autoencoder network \citep{Romero_2017}. Similar to conventional autoencoders, the goal is to compress and reconstruct the dimensionality of input data. This is achieved by optimizing an ansatz, $U(\bm{\theta})$, to compress the quantum state of the input and then restore it to a state closely approximating the original input. Figure~\ref{fig:QAE_before_swap} illustrates the structure of the target quantum circuit.
In the input $AB$ system (Figure~\ref{fig:QAE_before_swap}), $A$ corresponds to the latent state, where input information is compressed, and $B$ corresponds to the trash state. Additionally, $B'$ corresponds to the reference state, containing the same number of qubits as $B$. Successful compression and reconstruction are achieved if the input of $AB$ system matches the output of $AB'$ system after applying $U^\dagger(\bm{\theta})$. This means maximizing the fidelity between the input and output states. The input state of each qubit in the $AB$ system is $\ket{\psi_i}$, and the quantum state of the $B'$ system is $\ket{a}$. The density operator of the $AB'$ system, which includes the output state, is obtained by tracing out system $B$ from the entire system. This can be expressed by the following equation:
\begin{equation}
\rho_{AB'} = U_{AB'}^\dagger(\bm{\theta}) \mathrm{Tr}_B \big[ U_{AB}(\bm{\theta}) \big[ \ket{\psi_i} \bra{\psi_i}_{AB} \otimes \ket{a} \bra{a}_{B'} \big] U_{AB}^\dagger(\bm{\theta}) \big] U_{AB'}(\bm{\theta}). \label{eq:rho_AB'}
\end{equation}
The objective function $C(\bm{\theta})$, which must be maximized, can be defined as follows:
\begin{equation}
C(\bm{\theta}) = \sum_i p_i F \big(\ket{\psi_i}, \rho_{AB'} \big), \label{eq:C1}
\end{equation}
where $p_i$ is the probability that $\ket{\psi_i}$ is the output state.
However, determining the match between the quantum states of the $AB$ and the $AB'$ systems would typically require numerous qubit measurements. A swap test is performed on $B$ and $B'$ to address this challenge. The fidelity of the input and output states of the $AB$ system after swapping the quantum states of $B$ and $B'$ can be expressed as follows:
\begin{align}
F \big(\ket{\psi_i}, \rho_{AB} \big) &= F \big( \ket{\psi_i}, U_{AB}^\dagger \mathrm{Tr}_{B'} \big[ U_{AB'} \big[ \ket{\psi_i} \bra{\psi_i}_{AB'} \otimes \ket{a} \bra{a}_{B} \big] U_{AB'}^\dagger \big] U_{AB} \big) \nonumber \\
&= \bra{\psi_i} U_{AB}^\dagger \mathrm{Tr}_{B'} \big[ U_{AB'} \big[ \ket{\psi_i} \bra{\psi_i}_{AB'} \otimes \ket{a} \bra{a}_{B} \big] U_{AB'}^\dagger \big] U_{AB}  \ket{\psi_i}. \label{eq:F_AB}
\end{align}
The fidelity of the input and output states of the $AB'$ system before the swap can be expressed as follows:
\begin{equation}
F \big( U_{AB'} \ket{\psi_i}, \mathrm{Tr}_{B} \big[ U_{AB} \ket{\psi_i} \bra{\psi_i} U_{AB}^\dagger \otimes \ket{a} \bra{a}_{B'} \big] \big). \label{eq:F_AB'}
\end{equation}
After swapping $B$ and $B'$, the fidelity can be rewritten as follows, demonstrating equivalence with the fidelity of the $AB$ system.
\begin{align}
F &\big( U_{AB} \ket{\psi_i}, \mathrm{Tr}_{B'} \big[ U_{AB'} \ket{\psi_i} \bra{\psi_i} U_{AB'}^\dagger \otimes \ket{a} \bra{a}_{B} \big] \big) \nonumber \\
&= \bra{\psi_i} U_{AB}^\dagger \mathrm{Tr}_{B'} \big[ U_{AB'} \big[ \ket{\psi_i} \bra{\psi_i}_{AB'} \otimes \ket{a} \bra{a}_{B} \big] U_{AB'}^\dagger \big] U_{AB}  \ket{\psi_i} \nonumber \\
&= F \big(\ket{\psi_i}, \rho_{AB} \big) \label{eq:F_AB'_swap}
\end{align}
Therefore, swapping $B$ and $B'$ is equivalent to reconstructing the input. The swap test leverages the agreement between quantum states, which can be determined by measuring a single ancillary qubit, significantly reducing the measurement overhead compared to the pre-swap scenario. 

\begin{figure}[t]
\centering
\includegraphics[width=0.5\textwidth]{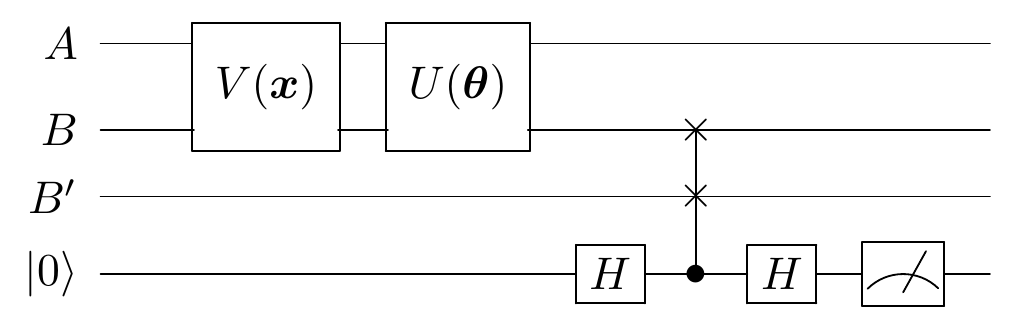}
\caption{Quantum autoencoder structure after conducting the swap test}\label{fig:QAE_with_swap}
\end{figure}

Figure~\ref{fig:QAE_with_swap} depicts the quantum circuit configured for learning image data with a QAE.
Image data are encoded into the quantum state by setting them as parameters in the circuit $V(\bm{x})$. The parameter $\bm{\theta}$ of the ansatz $U(\bm{\theta})$ is optimized using a classical algorithm to maximize the agreement between the $B$ and $B'$ states. The swap test is introduced to facilitate this. As mentioned previously, the swap test is an algorithm for determining the agreement of two quantum states. An auxiliary qubit is prepared for measurement; if the two quantum states match, the measurement result of the auxiliary qubit will be $\ket{0}$. Consequently, by defining an objective function that minimizes the probability of measuring $\ket {1}$, the parameter $\bm{\theta}$ is optimized to align the states of $B$ and $B'$.

Similar to classical autoencoders, QAEs have also been applied to tasks such as anomaly detection, data compression, and noise reduction \citep{Pepper_2019, Ngair_2022, Sakhnenko_2022, Hui_2022, Zhu_2023, Mok_2024}. For the task related to image data, results in \cite{Wang_2024} have shown that efficient compression may be achieved while preserving essential features. In several cases, classification problems have also been addressed. As with approaches using classical autoencoders, the features extracted by QAEs are used for classifying data \citep{Srikumar_2021, Mangini_2022}.

\section{Quantum autoencoders for image classification}

\begin{figure}[t]
\centering
\includegraphics[width=0.5\textwidth]{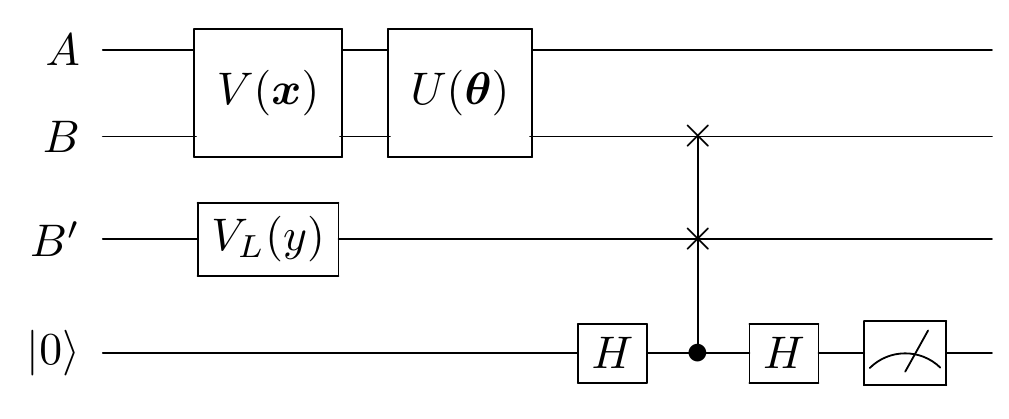}
\caption{Quantum autoencoder circuit for image-classification training. Class information is encoded in $V_L(y)$}\label{fig:QAE_classification_train}
\end{figure}

\begin{figure}[t]
\centering
\raisebox{2.6em}{
\begin{subfigure}{0.3\textwidth}
    \centering
    \includegraphics[width=\linewidth]{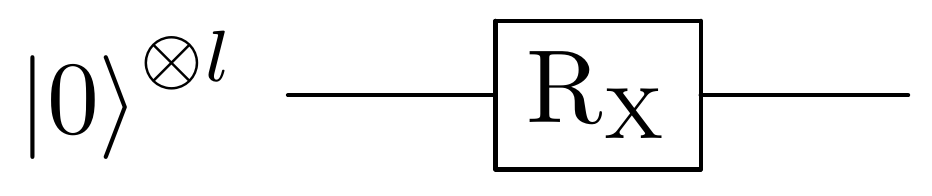}
    \caption{The definition of $V_L(y)$}
    \label{fig:ref_state_definition}
\end{subfigure}
}
\hspace{0.05\textwidth}
\begin{subfigure}{0.29\textwidth}
    \centering
    \includegraphics[width=\linewidth]{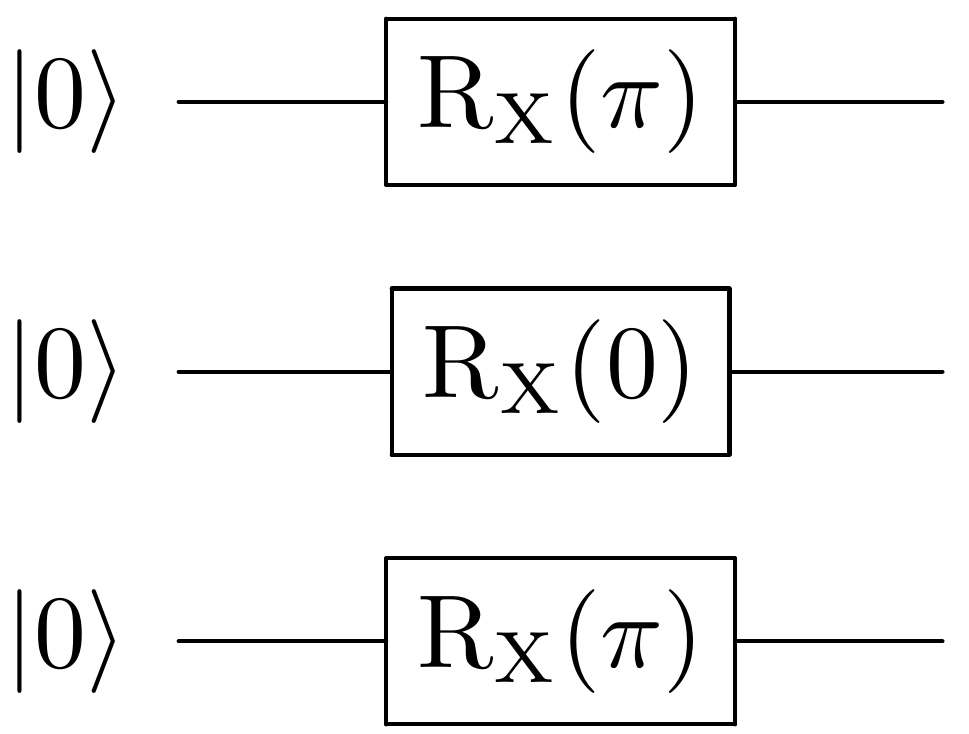}
    \caption{$V_L$ which represents the quantum state $\ket{101}$}
    \label{fig:ref_state_101}
\end{subfigure}
\caption{Circuit diagram for $V_L(y)$, used for encoding class information into the reference state $B'$}
\label{fig:reference_state_for_classification}
\end{figure}

This study proposes a novel quantum circuit for image-classification training using QAEs. Figure~\ref{fig:QAE_classification_train} provides an overview of the circuit.
The circuit $V(\bm{x})$ for encoding input image data, ansatz circuit $U(\bm{\theta})$, and swap test are based on the original QAE architecture. To adapt the QAE for image classification, the reference state $B'$ is set using the circuit $V_L(y)$. Figure~\ref{fig:ref_state_definition} illustrates the structure of $V_L(y)$.
In the $V_L(y)$ circuit, an RX gate is assigned to each qubit. The parameter for the RX gate corresponds to the rotation angle of the quantum state around the $x$-axis, thus changing the state of the qubit based on the parameter value. When a qubit is initialized to the state $\ket{0}$, and the RX gate parameter is set to $\pi$, the state flips to $\ket{1}$. This phenomenon is used to represent class information. If $B'$ comprises $l$ qubits, there are $2^l$ possible quantum states, allowing for assigning up to $2^l$ different classes. For example, consider the case where $l = 1$ and the state $\ket{0}$ represents class $y = 0$, whereas $\ket{1}$ represents class $y = 1$. If the image data input into $V(\bm{x})$ belongs to class 0, the state of $V_L(y=0)$ should remain $\ket{0}$. To achieve this, the RX gate parameter is set to zero, leaving the initial state unchanged. Conversely, if it belongs to class 1, the $V_L(y=1)$ state should be $\ket{1}$. This is achieved by setting the RX gate parameter to $\pi$, flipping the initial state. 
As a more advanced use of the trash state, the multiclass case is considered. When $l = 2$, up to four classes can be expressed, and when $l = 3$, up to eight classes can be expressed by the trash state $B'$. All representable quantum states, ranging from all qubits in the $\ket{0}$ state to all in the $\ket{1}$ state, are sequentially utilized. As in the case of $l=1$, when a qubit needs to be flipped to the $\ket{1}$ state, an RX gate is applied. For example, when $l=3$, the quantum state $\ket{101}$ represents class 5. To represent class 5, the RX gate parameters in the first and third qubits of $B'$ are set to $\pi$. This circuit structure is illustrated in Figure~\ref{fig:ref_state_101}.
Under these conditions, parameter optimization learning proceeds similarly to that in previous QAE studies \citep{Romero_2017}. The parameters of $U(\bm{\theta})$ are trained such that $B$ encodes class information based on the swap test results between $B$ and $B'$. 

\begin{figure}[t]
\centering
\includegraphics[width=0.5\textwidth]{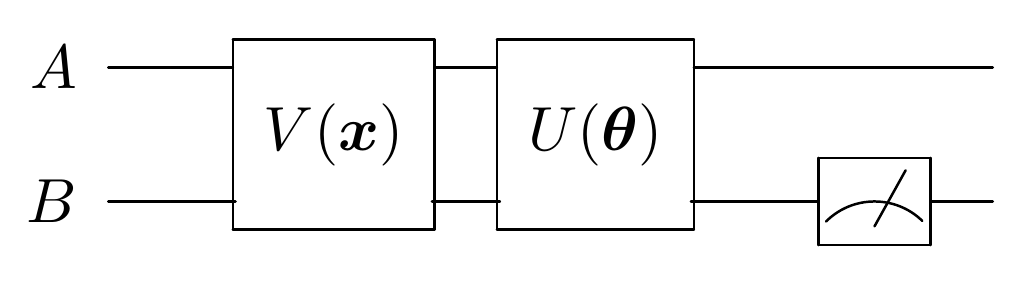}
\caption{Quantum autoencoder circuit for image-classification prediction}\label{fig:QAE_classification_test}
\end{figure}

Figure~\ref{fig:QAE_classification_test} shows the circuit used for classifying images using the trained $\bm{\theta}$.
The parameters of $V(\bm{x})$ are set using the information from the test image to be classified and the trained parameters are applied to $U(\bm{\theta})$. The circuit is then measured. Because $\bm{\theta}$ has been optimized using the swap test results between $B$ and $B'$ as the objective function, the class information of the image is output to $B$. The class of the test image can be predicted by comparing the qubit measurement results of the quantum state in $B$ with the correspondence between quantum states and class information defined in $B'$ during training. The predicted result is the class corresponding to the quantum state with the highest probability among the measurement results. In conventional QAE methods, $B$ is in a trash state and unused for tasks such as information recovery. However, the proposed method encodes label information into $B$, allowing the qubits in $B$ to be effectively used for image classification. 

Existing classification methods using QAE utilize features of data compressed into latent states \citep{Srikumar_2021, Mangini_2022}. Since these features are learned unsupervised without being associated with class information, it is necessary to prepare a separate classifier. The proposed method enables classification within the QAE algorithm by performing training that inputs class information into the trash state. This approach constitutes a supervised autoencoder and represents a methodology not found in conventional QAEs or classifications using classical autoencoders.
In current approaches to image classification using quantum machine learning, QCNNs are widely employed; however, as described in Section \ref{sec:QCNN}, they heavily depend on classical algorithms. In the proposed method, classical data are encoded into quantum states by quantum circuits, and these quantum states are analyzed through an ansatz. Therefore, the performance of the quantum circuit significantly affects the training performance. Although classical optimization algorithms are used to optimize the parameters of the ansatz, a major distinction from the QCNN approach is that quantum computation is responsible for most of the learning required for classification.

\section{Experiments}

\subsection{Experimental settings}\label{sec:experimental_settings}
To validate the proposed image-classification approach, experiments were conducted using three different datasets. The MNIST dataset consists of handwritten digit images. Fashion-MNIST is a dataset in the same format as MNIST, but containing images of clothing items such as T-shirts and trousers \citep{Xiao_2017}. Kuzushiji-MNIST (KMNIST), on the other hand, features images of Japanese cursive characters known as Kuzushiji \citep{Clanuwat_2018}. In this experiment, the conditions are restricted to grayscale image classification. Color images are more representative of realistic problem settings, and methods such as Multi-Channel Representation for Quantum Image (MCRQI) make it possible to encode them into quantum states \citep{Sun_2011}. However, in such cases, the number of qubits required for training increases compared to the grayscale setting. Under current quantum hardware capabilities, it is difficult to perform large-scale quantum computation with stability. Additionally, since this is the initial verification of QAE classification, the experiment aims to observe the results under a small-scale problem setting. In this experiment, subsets were created by randomly selecting images belonging to four classes (0–3) from each dataset. The training set comprised 500 images, whereas the test set comprised another 500. The dataset was constructed to contain a nearly equal number of samples for each class, mitigating the risk of class imbalance.

In this experiment, amplitude encoding is used as the image encoder $V(\bm{x})$ \citep{Rath_2024}. Amplitude encoding maps classical data $\bm{x}$ of dimension $2^n$ to the probability amplitudes of the $n$-qubits quantum state. The datasets used in this experiment, namely the MNIST dataset, the Fashion MNIST dataset, and the KMNIST dataset, are originally distributed in a size of $28 \times 28$. Because the dimension of $\bm{x}$ must be an integer when calculating $\log_{2}2^n$, datasets were resized before encoding. Although maintaining resolution as close as possible is desirable, the image size affects the scale of the $AB$ system. In this experiment, the image size was primarily set to $16 \times 16$, with variations applied in certain cases. Consequently, the $AB$ system comprised $n = 8$ qubits when using amplitude encoding.

Various effective methods for encoding classical image data into quantum states have been proposed, such as basis encoding, angle encoding, novel enhanced quantum representation (NEQR), and flexible representation of quantum images (FRQI) \citep{Rath_2024, Zhang_2013, Phuc_Q_2011}. However, these methods require a greater number of qubits compared to amplitude encoding. While FRQI is relatively efficient in terms of qubit usage, its practical implementation remains complex due to the need to execute quantum circuits during the encoding process. Amplitude encoding is sufficient to represent the information in the simple image datasets used in this experiment.


Given the four-class dataset, four distinct quantum states were necessary for classification; these states required a minimum of two qubits. The trash state $B$, which encodes class information, was set to $l = 3$ qubits. This setting allowed for a comparative experiment with an eight-class classification task using a QAE classification circuit of the same structure. Eight-class classification necessitates eight quantum states, representable with a minimum of three qubits. Maintaining the same number of qubits for the trash and reference states in the $AB$ system facilitated this comparison. The remaining $k = n - l$ qubits in $AB$ were used for the latent state $A$. The reference state $B'$ had the same number of qubits as $B$ and was, therefore, also set to $l$ qubits. An additional qubit was required for the swap test. Thus, $n + l + 1$ qubits were used in this QAE classification experiment.

The classical optimization method constrained optimization by linear approximation (COBYLA) was employed to optimize the parameter $\bm{\theta}$ of the ansatz $U(\bm{\theta})$. A key characteristic of COBYLA is that it does not require gradient information. Because current quantum hardware is highly susceptible to noise, gradient-based optimization using numerical differentiation can be unstable. COBYLA, being gradient-free, is expected to yield more stable results in QCL \citep{Bonet_Monroig_2023, Pellow_Jarman_2024}. However, this experiment utilized a quantum circuit simulator, which is noise-free compared to actual quantum hardware. The proposed method was designed with future implementation on real hardware in mind; therefore, results using COBYLA are presented herein. In this experiment, parameter optimization with COBYLA was primarily performed for 5000 epochs.

For qubit measurement during training and testing, the quantum circuit simulator provided by Qiskit was used \citep{Qiskit}. Specifically, we employed the the statevector simulator, which emulates the theoretical state of a quantum computer, computing the state vector after quantum circuit execution. Qiskit also offers the Qasm simulator, which probabilistically samples measurement results and access to gate-based quantum computers via the cloud. However, given the current state of quantum error correction and hardware, quantum measurements using these methods are highly susceptible to noise-induced errors. This experiment excluded the influence of device noise on computational accuracy, presenting results obtained under ideal conditions as an indicator of the potential of the proposed method. A seed value was set for all processes that would otherwise produce stochastic results, ensuring reproducibility.

\begin{figure}[H]
\centering
\begin{subfigure}{0.3\textwidth}
    \centering
    \includegraphics[width=\linewidth]{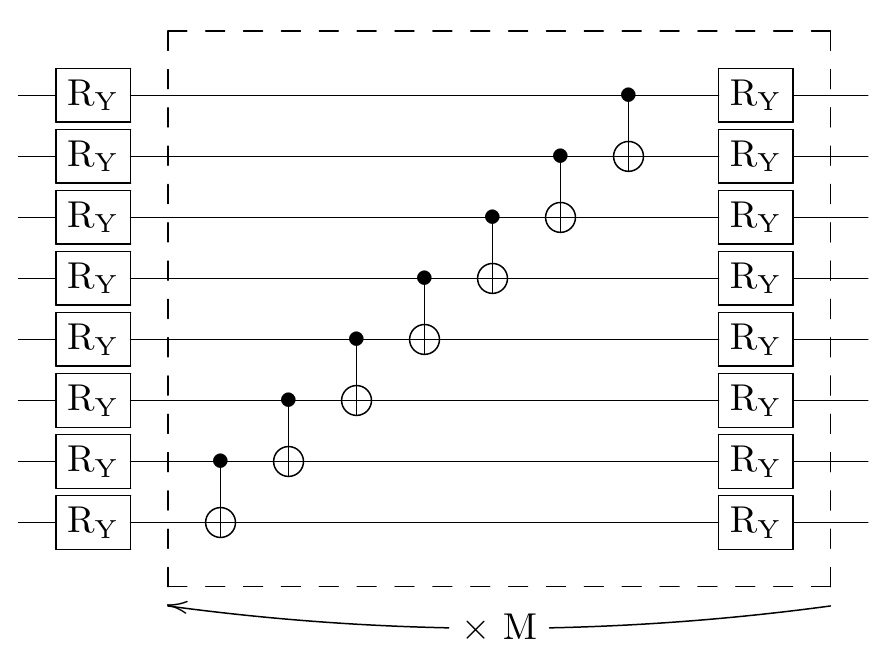}
    \caption{Circuit 1}
    \label{fig:circ1}
\end{subfigure}
\begin{subfigure}{0.29\textwidth}
    \centering
    \includegraphics[width=\linewidth]{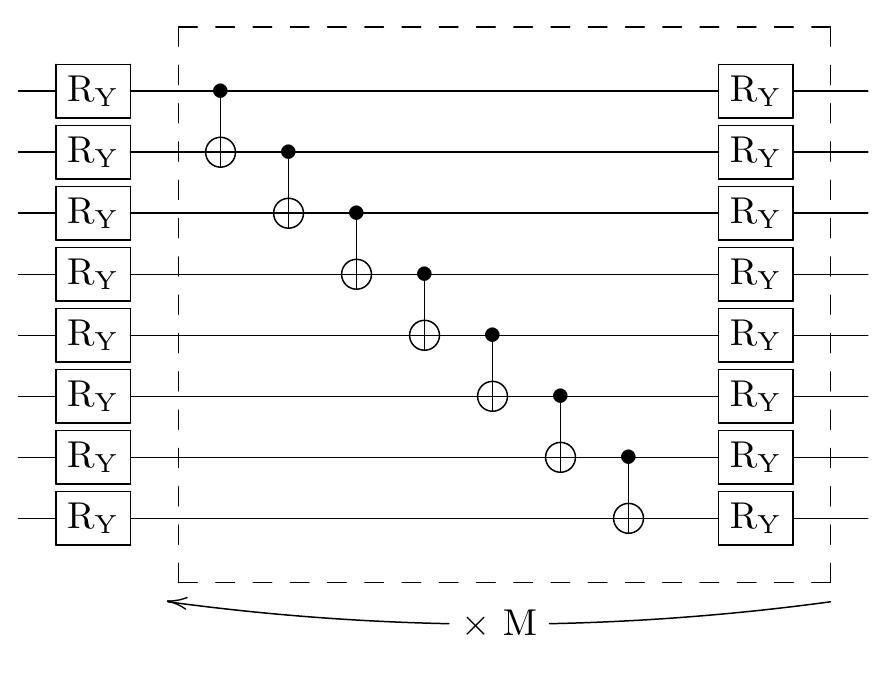}
    \caption{Circuit 2}
    \label{fig:circ2}
\end{subfigure}
\begin{subfigure}{0.32\textwidth}
    \centering
    \includegraphics[width=\linewidth]{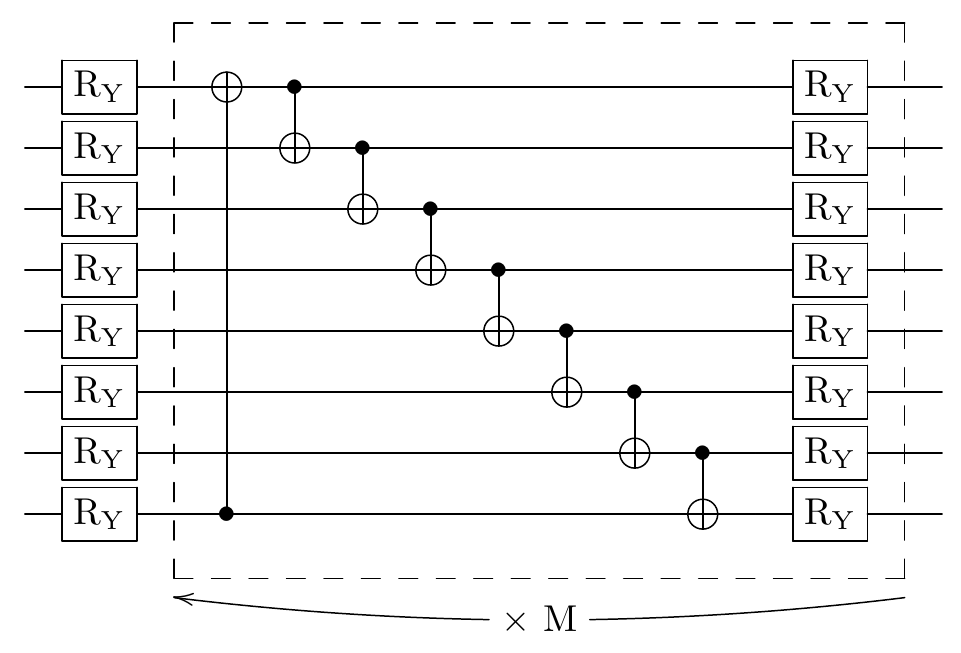}
    \caption{Circuit 3}
    \label{fig:circ3}
\end{subfigure}
\begin{subfigure}{1.0\textwidth}
    \centering
    \includegraphics[width=\linewidth]{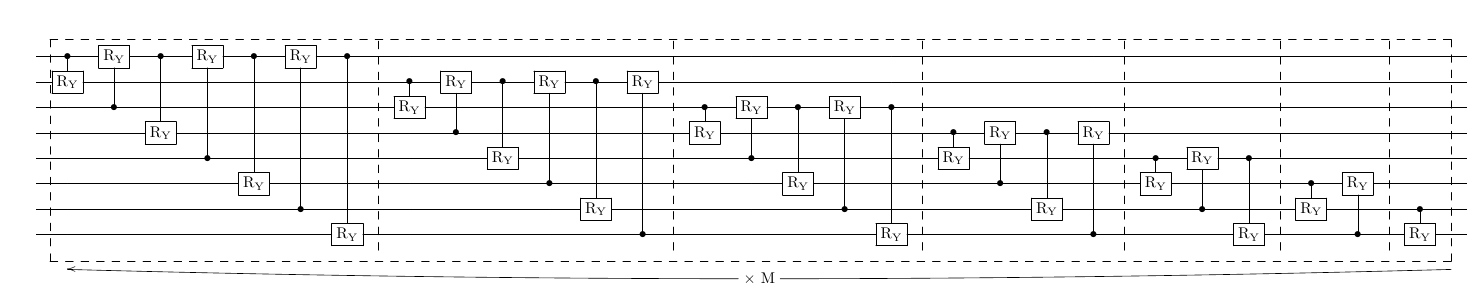}
    \caption{Circuit 4}
    \label{fig:circ4}
\end{subfigure}
\begin{subfigure}{1.0\textwidth}
    \centering
    \includegraphics[width=\linewidth]{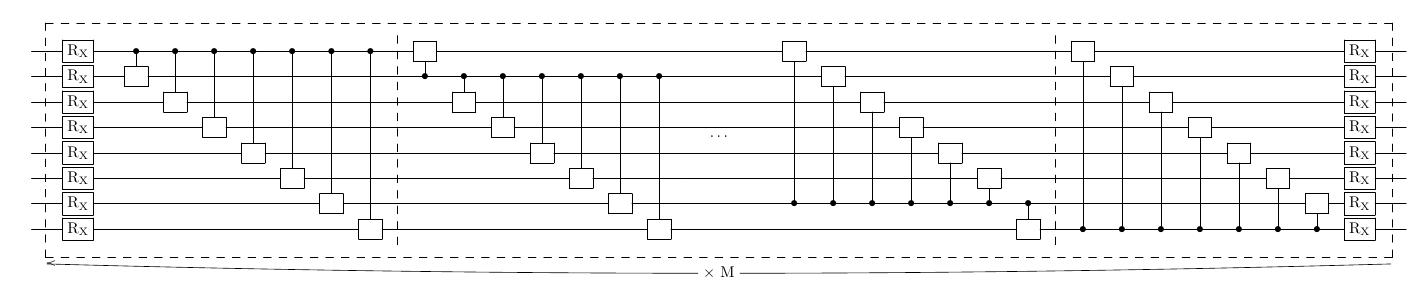}
    \caption{Circuit 5}
    \label{fig:circ5}
\end{subfigure}
\begin{subfigure}{0.6\textwidth}
    \centering
    \includegraphics[width=\linewidth]{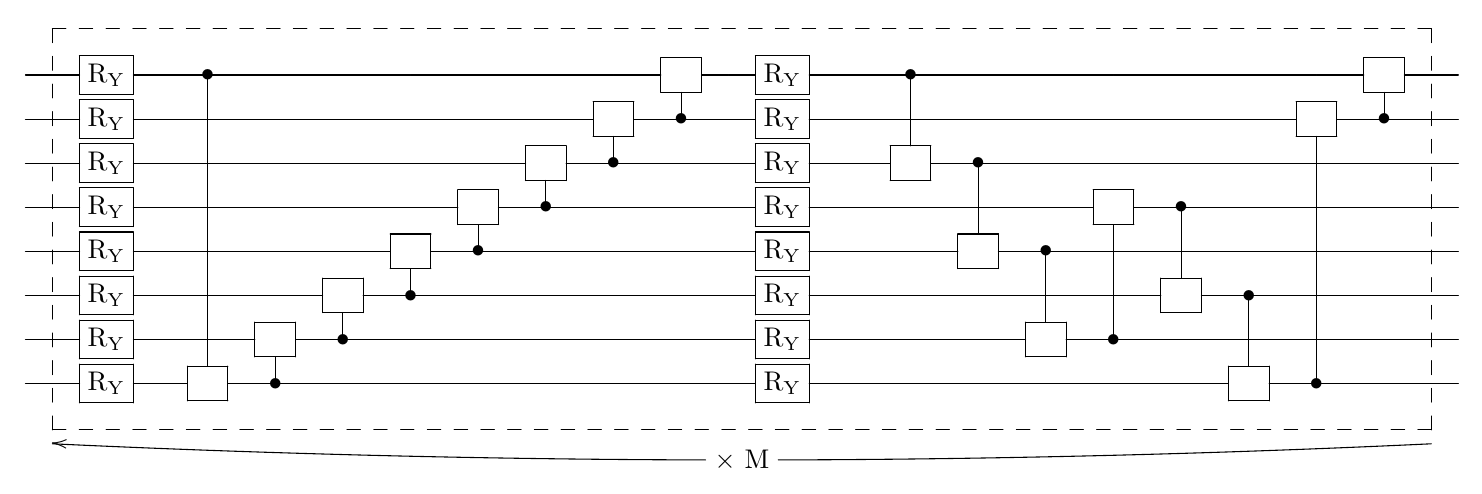}
    \caption{Circuit 6}
    \label{fig:circ6}
\end{subfigure}
\begin{subfigure}{0.35\textwidth}
    \centering
    \includegraphics[width=\linewidth]{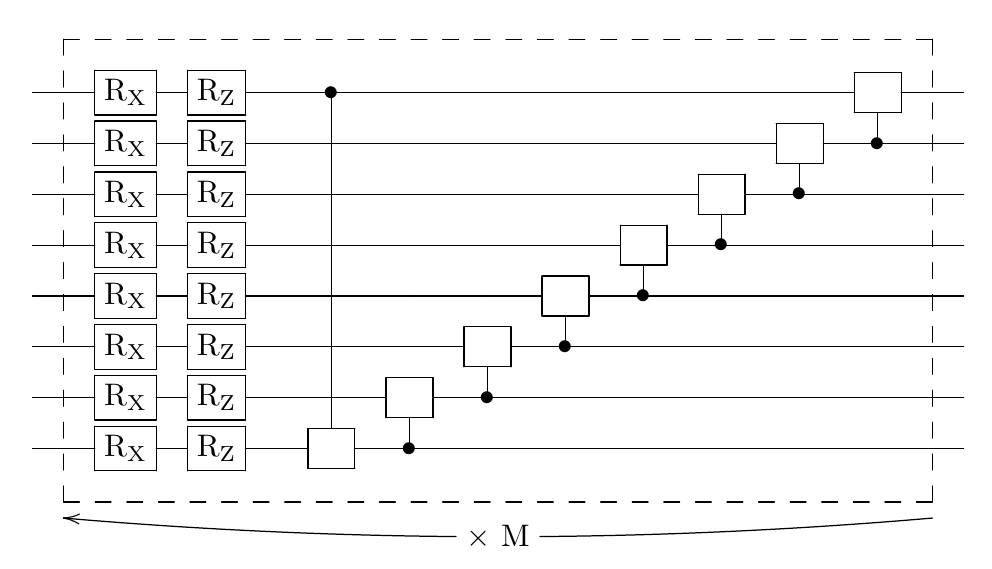}
    \caption{Circuit 7}
    \label{fig:circ7}
\end{subfigure}
\caption{Quantum circuit structures for different parameterized ansatzes (Circuits 1–7). Circuits 1–3 have real-amplitude circuit variations, whereas Circuits 4–7 are based on previous studies. $M$ represents the number of repetitions in each ansatz. The blank regions within the circuits indicate positions where arbitrary quantum gates are configured during the experiments}
\label{fig:circ_1_to_5}
\end{figure}

\subsection{Ansatz}
The construction of the ansatz $U(\bm{\theta})$ in the QAE classification circuit primarily employs several types of ansatzes frequently used for training conventional QAEs. Figure~\ref{fig:circ_1_to_5} illustrates the structure of each ansatz. The blank regions within the circuits indicate positions where arbitrary quantum gates are configured during the experiments. Each ansatz includes a section that is repeated multiple times, as indicated in the ansatz schematic. The number of repetitions is denoted as $M$.

The first ansatz considered is the real-amplitude circuit, wherein classical information is initially encoded through rotations using RY gates \citep{Alessandro_2022}. Subsequently, adjacent qubits are entangled via CNOT gates, followed by additional rotations using RY gates to adjust the quantum state. This entanglement sequence with CNOT gates and rotation with RY gates is repeated $M$ times. Several patterns exist for applying entanglement via CNOT gates.
One such pattern is the reverse linear pattern, which begins by applying a CNOT gate with the control and target qubits being the second-to-last and last qubits, respectively. The control qubit is then shifted sequentially, one by one, up to the first qubit, with CNOT gates applied at each step. This configuration is presented as Circuit 1 in Figure~\ref{fig:circ1}. Conversely, the linear pattern applies CNOT gates sequentially, with the control qubits starting from the first and moving down to the lower qubits. This configuration is presented as Circuit 2 in Figure~\ref{fig:circ2}. A variation similar to the linear pattern, the circular pattern, adds a CNOT gate to control the first qubit from the last qubit before establishing entanglement among adjacent qubits. This creates a circular connection among qubits, enhancing entanglement, and is shown as Circuit 3 in Figure~\ref{fig:circ3}. The real-amplitude circuit represents a simple circuit structure; its effectiveness in image-classification tasks, along with variations owing to different entanglement patterns, was examined in the experiments.

Circuits 4 and 5 adopt the ansatz structures proposed in the original QAE. Circuit 4 entangles all possible qubit pairings by applying Controlled-RY gates. The application of Controlled-RY gates to all qubit pairs constitutes one iteration, repeated $M$ times. Figure~\ref{fig:circ4} presents the structure of Circuit 4.
Circuit 5 is designed to incorporate all possible controlled rotations for each qubit. Initially, each qubit undergoes a rotation using an RX gate. Subsequently, the first qubit controls all other qubits using Controlled-RY gates. This process is performed sequentially for each qubit in the ansatz, moving from the first to the last. Once the last qubit has completed its control operations, each qubit undergoes another RX rotation, similar to that in the initial step. This entire sequence constitutes one iteration and is repeated $M$ times. The structure of Circuit 5 is presented in Figure~\ref{fig:circ5}. Due to its lengthy configuration, parts of the circuit are omitted in the figure using an ellipsis for clarity.
Circuits 4 and 5 have been used in previous research to compress the ground states of the Hubbard model and molecular Hamiltonians \citep{Romero_2017}. The present study investigated their applicability for training in image-classification tasks.

Circuits 6 and 7 are considered effective for classification tasks \citep{Schuld_2020}. Figure~\ref{fig:circ6} and \ref{fig:circ7} presents their structures. They generate strong entanglement between qubits through the arrangement of controlled gates. Input states can be effectively mapped into a high-dimensional Hilbert space, enabling representations that are more distinguishable for classification. However, these circuits are designed to obtain a binary class prediction result by measuring a single qubit. This study verifies whether it may also perform effectively in multiclass classification tasks.

\subsection{Other machine learning methods for comparison}

\subsubsection{Nonnegative/binary matrix factorization}
NBMF was originally proposed as a generative model trained using a quantum annealing machine \citep{O_Malley_2018}. It is an algorithm that optimizes features necessary for reconstructing the original data by factorizing a matrix into nonnegative real-valued basis and binary coefficient matrices. Optimizing the elements of the binary matrix can be viewed as a combinatorial optimization problem suitable for annealing methods. Previous studies have proposed NBMF as a multiclass image-classification model that takes a matrix comprising image data and corresponding class information as the input \citep{Asaoka_2023}. NBMF decomposition of this input yields a basis matrix that extracts features associating the image data with the class information. This matrix enables class prediction for test data.
The training process in NBMF for image classification leverages nonnegativity constraints and a binary combinatorial optimization procedure. This allows for effective feature extraction in the early stages of learning, even with small datasets and limited features, resulting in high accuracy. Leveraging quantum optimization to retain only necessary information is expected to further improve accuracy, even with less data.

Since NBMF can serve as an image classification model that leverages quantum annealing, a comparative experiment was conducted to evaluate its performance against the proposed method, which is based on gate-based quantum circuits. The feature dimension in NBMF is equivalent to the coefficient matrix's dimension and the number of columns in the basis matrix. This feature dimension can be considered equivalent to the information content in the classical space of the latent state in QAE classification. A comparative experiment was conducted with the feature dimension set to $2^k = 32$, decomposing the dataset under conditions identical to those in the QAE classification evaluation.

Owing to the presence of zero elements in the binary coefficient matrix, NBMF inherently exhibits high sparsity in its trainable parameters. Even in its original formulation, approximately 74\% of the parameters are sparse. In addition to this baseline, results were also obtained under experimental conditions designed to more closely align the number of parameters with those used in QAE classification. By introducing a constraint term into the update rule for the binary coefficient matrix, the sparsity was further increased, resulting in a reduction in the total number of trainable parameters.

\subsubsection{Neural network}
In a fully connected neural network (FCNN), neurons are connected between consecutive layers. The network learns features by propagating information from the input to the output layer. Image data are provided as input and transformed through the hidden layer, and the predicted class is output. The network is trained using the multiclass cross-entropy loss between the output layer and the true class labels as the objective function. Parameters are updated to minimize this objective function, thereby minimizing the difference between the FCNN's output and the actual class labels. 

Because the FCNN can be configured to have the same structure as NBMF by aligning the experimental conditions, it was employed as a comparative method. An FCNN can be structurally mapped to NBMF's feature extraction process through a three-layer architecture. The nodes in the hidden layer correspond to the column vectors of the coefficient matrix in NBMF, while the edge weights correspond to the components of the basis matrix. The number of nodes in the FCNN’s hidden layer corresponds to the feature dimension in NBMF, which is equal to the dimension of the coefficient matrix and the number of columns in the basis matrix.

As one of the experimental settings, a three-layer FCNN (input, hidden, and output) was used with $2^k = 32$ nodes in the hidden layer. It was implemented as a simple multilayer perceptron with a single hidden layer, employing the sigmoid function as the activation function. This configuration was chosen to match the dimensionality of extracted features with that used in QAE classification and NBMF. The Adam optimizer was employed for parameter optimization \citep{Kingma_2015}.

Inspired by the sparsity of the coefficient matrix in NBMF, additional experiments were conducted using sparse neural networks (SNNs) in which the trainable parameters were sparsified. The structure of the SNN is based on that of the FCNN. Two types of SNNs were employed: one with a sparsity level of 74\% to match that of NBMF, and the other configured to match the number of parameters used in QAE classification.

\subsubsection{Quantum convolutional neural network}\label{sec:QCNN}
As QCNN is frequently adopted as an approach to image classification in quantum machine learning, it is necessary to compare its performance with the proposed method. However, the problem size that can be realistically executed within a reasonable time using a quantum simulator is limited. In this study, a QCNN architecture was employed in which quantum convolution is applied only to a portion of the network \citep{Henderson_2019, Matic_2022}. The implementation leveraged publicly available code \citep{QC-CNN}. The network comprises a quantum convolutional layer that performs convolution operations via quantum circuits, followed by a classical fully connected layer. In the quantum convolutional layer, image patches are extracted using a 2×2 kernel with a stride of 2. A quantum circuit is then applied to each patch to extract features. The circuit consists of four qubits, where each qubit encodes input values using Hadamard and RY gates. Subsequently, each layer applies controlled-RZ gates between adjacent qubits and local RY gates. Measurements are performed in the Z basis for each qubit, and the expectation values are used as output. A total of 64 four-dimensional outputs obtained from the entire image serve as the quantum feature vector, which is then passed to the fully connected layer. The fully connected layer generates four-class classification outputs via a 32-dimensional hidden layer. The Adam optimizer is used to optimize parameters in both the quantum circuit and the fully connected network.

Due to the incorporation of convolutional processing, QCNN cannot be evaluated under the same criteria as NBMF and FCNN. Its performance is compared independently with that of QAE classification. Additionally, since QCNN training is time-consuming and convergence was confirmed, the comparison is based on the results obtained after 1000 training epochs. The dataset used in this experiment is identical to that employed in the other experiments.

\section{Results}

Image-classification training using the QAE classification quantum circuit was conducted using multiple ansatz configurations. Table~\ref{tab:ansatz_accuracy} presents the ansatz structure for each pattern, along with the classification accuracy when performing class prediction on test images using the optimized parameters obtained during training.
\begin{table*}[t]
\caption{Performance comparison of various quantum circuit ansatzes  (Circuits 1–7) across multiple datasets. Each circuit is characterized by its type of arbitrary quantum gate (RY, RZ, RX, or X), the number of repetitions ($M$), and the total number of trainable parameters. Test accuracies are reported for three datasets: MNIST, Fashion MNIST, and KMNIST. The final column provides the average accuracy across these datasets. In the records of accuracy for each column, the highest value is highlighted in bold}\label{tab:ansatz_accuracy}
\centering
\resizebox{\textwidth}{!}{%
\begin{tabular}{@{}cccccccc@{}}
\toprule
Ansatz & Arbitrary quantum gate & $M$ & Parameters & MNIST & Fashion MNIST & KMNIST & Average \\
\midrule
Circuit 1 & -- & 20 & 168 & 80.6\% & 67.2\% & \textbf{69.6\%} & 72.5\% \\
Circuit 2 & -- & 20 & 168 & 87.6\% & 58.8\% & 69.4\% & 71.9\% \\
Circuit 3 & -- & 20 & 168 & \textbf{90.4\%} & 67.2\% & 65.4\% & \textbf{74.3\%} \\
Circuit 4 & -- & 6 & 168 & 82.0\% & \textbf{69.6\%} & 64.8\% & 72.1\% \\
Circuit 5 & RY & 3 & 216 & 73.6\% & 68.2\% & 66.8\% & 69.5\% \\
Circuit 5 & X & 11 & 176 & 24.6\% & 25.2\% & 16.4\% & 22.1\% \\
Circuit 6 & RZ & 5 & 160 & 79.6\% & 49.8\% & 48.4\% & 59.3\% \\
Circuit 6 & RX & 5 & 160 & 86.2\% & 49.6\% & 61.2\% & 65.7\% \\
Circuit 6 & X & 10 & 160 & 75.8\% & 67.6\% & 68.0\% & 70.5\% \\
Circuit 7 & RZ & 7 & 168 & 67.4\% & 50.4\% & 50.4\% & 56.1\% \\
Circuit 7 & RX & 7 & 168 & 73.6\% & 51.2\% & 49.8\% & 58.2\% \\
\botrule
\end{tabular}%
}
\end{table*}
Table~\ref{tab:gate_num}  presents the number of quantum gates included in each ansatz, categorized into CNOT gates, rotation gates, and controlled rotation gates.
\begin{table*}[t]
\caption{Gate composition analysis of different ansatz structures.
Summarization of the number of gates used in each ansatz circuit, categorized by gate type: CNOT gates, rotation gates, and controlled rotation gates}
\centering
\resizebox{\textwidth}{!}{%
\label{tab:gate_num}
\begin{tabular}{@{}cccccccc@{}}
\toprule
Ansatz & Arbitrary quantum gate & $M$ & CNOT gates & Rotation gates & Controlled rotation gates & Total \\
\midrule
Circuit 1 & --  & 20 & 140 & 168 & 0   & 308 \\
Circuit 2 & --  & 20 & 140 & 168 & 0   & 308 \\
Circuit 3 & --  & 20 & 160 & 168 & 0   & 328 \\
Circuit 4 & --  & 6  & 0   & 0   & 168 & 168 \\
Circuit 5 & RY  & 3  & 0   & 48  & 168 & 216 \\
Circuit 5 & X & 11 & 616 & 176 & 0   & 792 \\
Circuit 6 & RZ  & 5  & 0   & 80  & 80  & 160 \\
Circuit 6 & RX  & 5  & 0   & 80  & 80  & 160 \\
Circuit 6 & X & 10 & 160 & 160 & 0   & 320 \\
Circuit 7 & RZ  & 7  & 0   & 112 & 56  & 168 \\
Circuit 7 & RX  & 7  & 0   & 112 & 56  & 168 \\
\bottomrule
\end{tabular}%
}
\end{table*}
The rotation and controlled rotation gates contain trainable parameters, with their rotation angles optimized during the learning process. For ansatzes in which arbitrary quantum gates are assigned, the total number of gates is summarized for each type of gate configured in the experiment.

\begin{figure}[t]
\centering
\begin{subfigure}{0.45\textwidth}
    \centering
    \includegraphics[width=\linewidth]{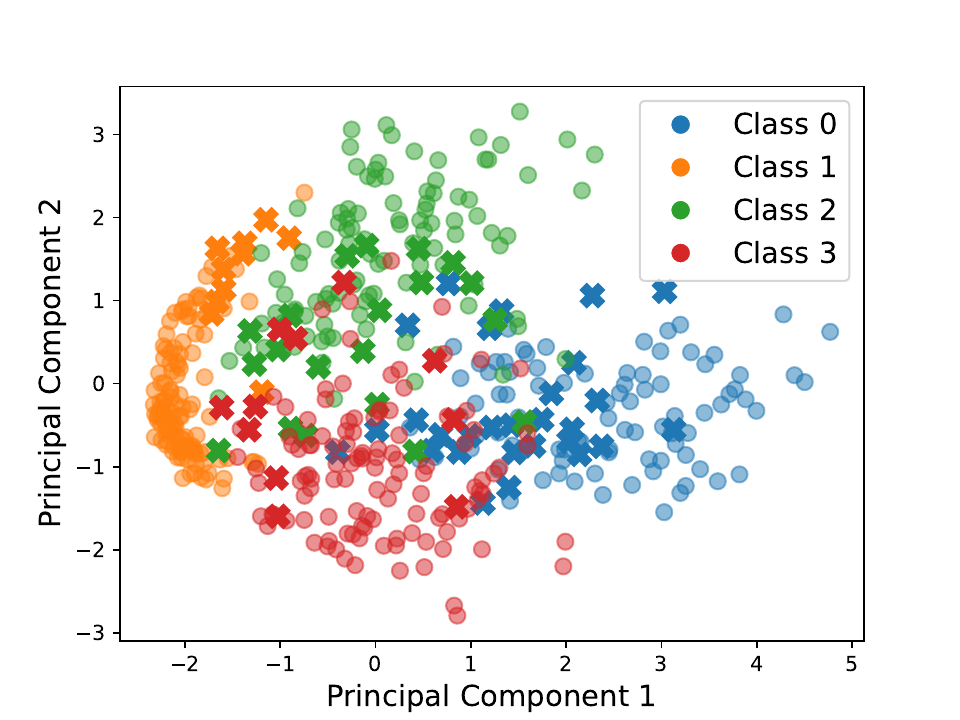}
    \caption{Circuit 6 with CRX gate}
    \label{fig:pca_circ1}
\end{subfigure}
\begin{subfigure}{0.45\textwidth}
    \centering
    \includegraphics[width=\linewidth]{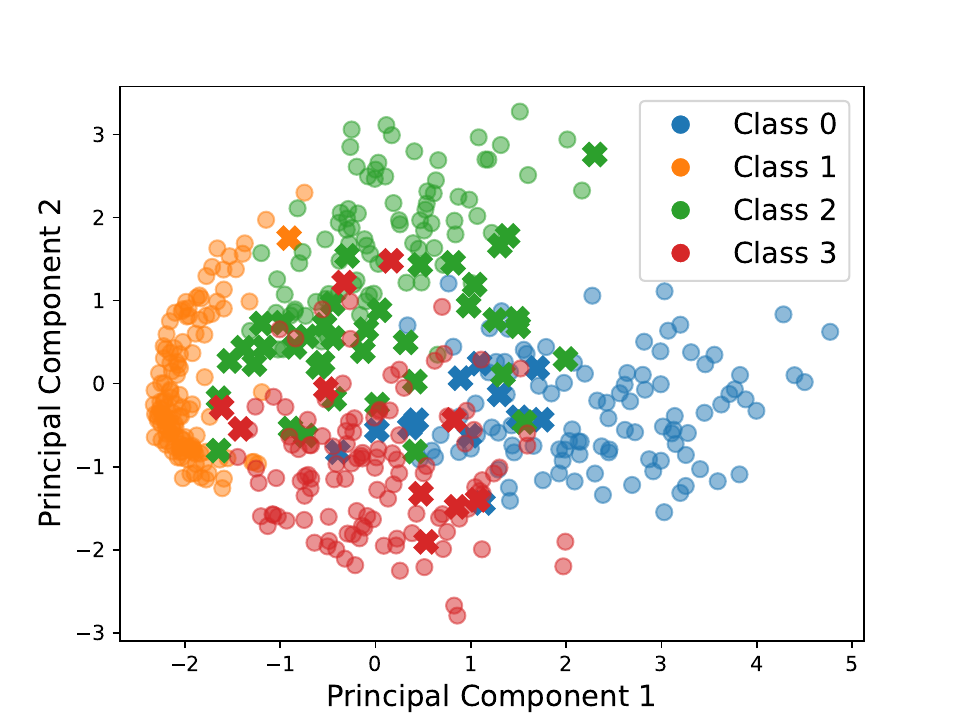}
    \caption{Circuit 2}
    \label{fig:pca_circ2}
\end{subfigure}
\vspace{10pt}
\begin{subfigure}{0.45\textwidth}
    \centering
    \includegraphics[width=\linewidth]{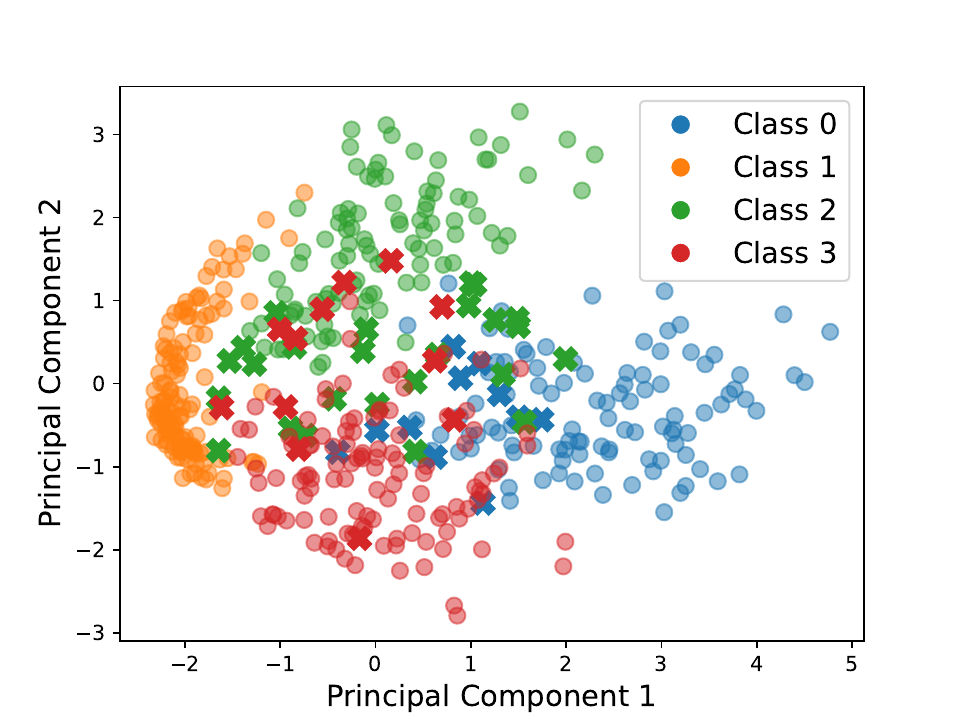}
    \caption{Circuit 3}
    \label{fig:pca_circ3}
\end{subfigure}
\caption{Principal component analysis (PCA) distributions for selected ansatzes: Circuit 6, 2, and 3. The ansatzes are arranged in increasing order of their classification accuracy on the MNIST dataset. Colors represent classes: Class 0 (blue), Class 1 (orange), Class 2 (green), and Class 3 (red). Circles indicate correct classification, whereas crosses indicate misclassification}
\label{fig:pca}
\end{figure}

The trends in classification accuracy differ depending on the dataset. The MNIST dataset achieved the highest classification accuracy among the datasets. Although the ansatz that achieved the highest accuracy varies across the three datasets, Circuit 3 consistently demonstrates the highest average accuracy.
To understand this result, we first examine the relationship between the classification results for each ansatz and the characteristics of the test data. Principal component analysis (PCA) was applied to reduce the dimensionality of the test image data to a two-dimensional feature vector representation. This study focuses on the results of the top three ansatz structures that achieved the highest classification accuracy on the MNIST dataset. Figure~\ref{fig:pca} plots the data points for each ansatz structure, class information, and classification results. 
The PCA plots are arranged in ascending order of test accuracy, specifically in the order of Circuit 6, 2, and 3. The color of each data point represents its class, as shown in the legend. Circles indicate correct classification, whereas crosses signify misclassification. For Circuit 3, most misclassified data points lie near the class boundaries in the feature space. In contrast, for other ansatz structures, misclassifications occur not only at class boundaries but also for data points with seemingly distinct class-specific features.

In this experiment, class prediction involved inputting quantum states created from test data into the trained ansatz and measuring the qubits of the trash state. The quantum state of the trash state with the highest prediction probability was adopted as the class prediction. The closer this highest probability was to 1, the higher the confidence in the class prediction. Using Circuit 3 as the ansatz structure is expected to yield predictions with higher confidence than other structures. Logarithmic loss is used to quantitatively evaluate this. Logarithmic loss, $L$, serves as a metric for assessing how well predicted probabilities align with the correct class and is defined by the following equation:
\begin{equation}
L = -\frac{1}{N} \sum_{i=1}^N \sum_{k=1}^C y_{ik} \log(p_{ik}),
\label{eq:log_loss}
\end{equation}
where $N$ represents the number of test data points and $C$ denotes the number of classes. The variable $y_{ik}$ is set to 1 if test data $i$ belong to class $k$, and zero otherwise. $p_{ik}$ represents the predicted probability that test data $i$ belong to class $k$. If the test data belong to class $k$ but are predicted with low confidence, $\log p_{ik}$ takes a large negative value. Frequent occurrences of such cases lead to an increased $L$. Conversely, correct predictions with high confidence result in a small $L$.
Table~\ref{tab:log_loss} presents the average logarithmic loss results obtained from QAE classification across multiple datasets using each ansatz. Although the average logarithmic loss and average accuracy for each ansatz do not exhibit a perfectly corresponding relationship, it can be observed that higher classification accuracy tends to be associated with lower logarithmic loss.
\begin{table*}[t]
\caption{Average logarithmic loss and test accuracy of various quantum circuit ansatzes, computed as the mean over three datasets: Digit MNIST, Fashion MNIST, and KMNIST. Each ansatz is defined by its quantum gate type and structure}
\label{tab:log_loss}
\centering
\resizebox{\textwidth}{!}{%
\begin{tabular}{@{}cccc@{}}
\toprule
Ansatz & Arbitrary quantum gate & Average test accuracy & Average logarithmic loss \\
\midrule
Circuit 1 & --   & 72.5\% & 0.9545 \\
Circuit 2 & --   & 71.9\% & 0.9579 \\
Circuit 3 & --   & 74.3\% & 0.9558 \\
Circuit 4 & --   & 72.1\% & 0.9875 \\
Circuit 5 & RY   & 69.5\% & 1.0270 \\
Circuit 5 & X & 22.1\% & 1.4272 \\
Circuit 6 & RZ   & 59.3\% & 1.1502 \\
Circuit 6 & RX   & 65.7\% & 1.1228 \\
Circuit 6 & X & 70.5\% & 0.9643 \\
Circuit 7 & RZ   & 56.1\% & 1.0598 \\
Circuit 7 & RX   & 58.2\% & 1.0103 \\
\bottomrule
\end{tabular}%
}
\end{table*}

To investigate the impact of different quantum-gate combinations, we analyzed the expressibility of each ansatz—a numerical metric quantifying the range of quantum states a circuit can represent \citep{Sim_2019, Hubregtsen_2021}. In principle, a highly expressive quantum circuit can produce any quantum state with nearly equal probability. This ideal distribution is approximated by the ensemble of Haar-random states, a uniform distribution over all possible quantum states. Therefore, expressibility is evaluated by calculating the Kullback–Leibler (KL) divergence between the distribution of output states generated by the ansatz and that of pure states under the Haar measure. Thus expressibility is denoted by $D_{KL}$. Lower $D_{KL}$, approaching zero, indicates higher expressibility, meaning the circuit can access a broader region of the Hilbert space. Because prediction confidence varies with the quantum-gate configuration, we hypothesized that the diversity of representable quantum states also differs between ansatzes. Consequently, we measured $D_{KL}$ of each ansatz. 
The implementation leveraged publicly available code \citep{Expr}. For this measurement, 2000 patterns of randomly generated, normalized real values were used as input. This input data size matched the image dimensions used in the experiment ($16 \times 16$), and these values were converted to quantum states via amplitude encoding. These encoded states were then input into each ansatz, using the parameters randomly generated from 0 to $2\pi$. 
$D_{KL}$ was then calculated from the resulting output states. Table~\ref{tab:exprs} presents the results, comparing $D_{KL}$ across the different ansatz structures. The circuits achieving higher accuracy tend to exhibit lower $D_{KL}$ values.
\begin{table*}[h]
\caption{$D_{KL}$ of different quantum circuit ansatzes. Lower $D_{KL}$ values indicate a broader range of representable quantum states. Average test accuracy computed as the mean over three datasets: Digit MNIST, Fashion MNIST, and KMNIST}\label{tab:exprs}%
\centering
\begin{tabular}{@{}cccc@{}}
\toprule
Ansatz & Arbitrary quantum gate & Average test accuracy & $D_{KL}$ \\
\midrule
Circuit 1 & -- & 72.5\% & 1.417831 \\
Circuit 2 & -- & 71.9\% & 1.423058 \\
Circuit 3 & -- & 74.3\% & 1.409319 \\
Circuit 4 & -- & 72.1\% & 1.679094 \\
Circuit 5 & RY & 69.5\% & 1.545853 \\
Circuit 5 & X & 22.1\% & 1.676397 \\
Circuit 6 & RZ & 59.3\% & 1.513340 \\
Circuit 6 & RX & 65.7\% & 1.532844 \\
Circuit 6 & X & 70.5\% & 1.438955 \\
Circuit 7 & RZ & 56.1\% & 1.466969 \\
Circuit 7 & RX & 58.2\% & 1.479359 \\
\botrule
\end{tabular}
\end{table*}

Table~\ref{tab:ansatz_8class} presents the results of applying Circuit 3, which achieved the highest average accuracy in Table~\ref{tab:ansatz_accuracy}, to an eight-class image-classification task of the MNIST dataset.
\begin{table}[t]
\caption{Test accuracies of QAE classification using Circuit 3 under various experimental conditions—number of classes, repetitions ($M$), parameters, training data size, image resolution, and latent qubit count—for classifying the MNIST dataset}\label{tab:ansatz_8class}%
\begin{tabular}{@{}ccccccc@{}}
\toprule
Classes & $M$ & Parameters & Training data & Image size & Latent qubits & Test accuracy \\
\midrule
4 & 20 & 168 & 500 & $16\times16$ & 5 & 90.4\% \\
8 & 20 & 168 & 500 & $16\times16$ & 5 & 49.4\% \\
8 & 30 & 248 & 500 & $16\times16$ & 5 & 55.2\% \\
8 & 20 & 168 & 800 & $16\times16$ & 5 & 63.6\% \\
8 & 20 & 210 & 500 & $32\times32$ & 7 & 56.4\% \\
\botrule
\end{tabular}
\end{table}
The initial conditions for the QAE classification circuit, including the training and test set sizes and configurations, remain unchanged, except that new data from classes four to seven are introduced into the dataset. As expected, the test accuracy decreases significantly. To improve accuracy, three approaches were implemented:
\begin{enumerate}
\item Increasing the number of ansatz repetitions from 20 to 30, thereby increasing the number of optimized parameters.
\item Expanding the training dataset from 500 to 800 images.
\item Expanding the input image size from $16 \times 16$ to $32 \times 32$.
\end{enumerate}
All modifications improved the test accuracy compared to the initial eight-class results; however, the accuracy did not reach the level achieved in the four-class classification task. In the third modification, the input image resolution was increased, thereby increasing the number of qubits in the latent state. The previous input image size of $16 \times 16$ was expanded to $32 \times 32$. Because $32 \times 32 = 2^{10}$, the number of qubits in the $AB$ quantum system increased to ten. With the trash state remaining at $l = 3$ qubits, the latent state could be expanded to $k = 7$ qubits. This expansion was expected to capture more class-specific features owing to the increased dimensionality of the latent space. However, there was no significant improvement in test accuracy away from expectations.

\begin{figure}[t]
\centering
\includegraphics[width=0.8\textwidth]{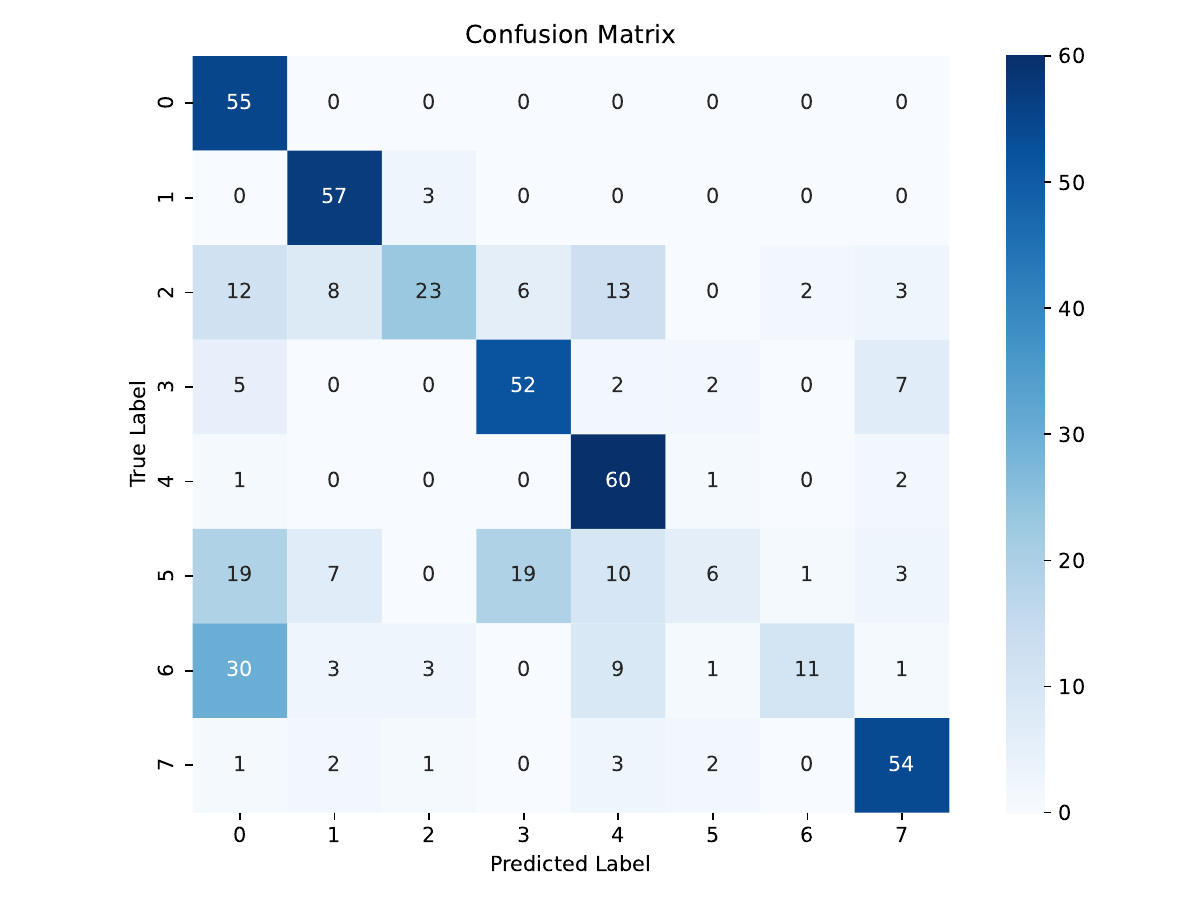}
\caption{Confusion matrix for the eight-class classification task of the MNIST dataset, showing predicted versus true labels. The matrix provides a detailed breakdown of correct and misclassified instances, wherein darker blue boxes indicate more samples in that cell. Diagonal elements represent correct classifications, whereas off-diagonal elements indicate misclassifications}\label{fig:prediction_heatmap}
\end{figure}

Figure~\ref{fig:prediction_heatmap} depicts the confusion matrix for the eight-class classification task after increasing the number of training samples (resulting in a test accuracy of 63.6\%).
The vertical axis represents the true class of the test data, whereas the horizontal axis indicates the predicted class output by the QAE classification model. Each cell's value represents the number of samples belonging to the true class (vertical axis) that were classified as the predicted class (horizontal axis).

A comparative analysis of four-class classification was conducted between QAE classification and conventional machine-learning methods, including NBMF, FCNN, and SNN. The same training and test datasets used in the QAE classification experiments were utilized. Parameter optimization was performed for 5000 epochs, similar to that in QAE classification. Table~\ref{tab:other_methods} presents the test accuracy for each dataset, achieved by the models trained using each method, corresponding to their respective model sparsity levels and the number of trainable parameters. Given that the dimensionality of the latent space in QAE classification is $k = 5$, the feature dimension in all other classical methods was set to $2^k = 32$. Sparsity refers to the proportion of parameters within a model that are effectively inactive in learning—specifically, those that take a value of zero and thus do not contribute to the optimization process. In the experiments where the number of parameters in NBMF and SNN was adjusted to approximate that of QAE classification, sparsity was introduced by imposing constraints on the model. Since the number of parameters was not explicitly specified, the exact sparsity levels and parameter counts varied slightly across experiments.
\begin{table*}[t]
\caption{Comparison of different classification methods with respect to parameter count, sparsity, and test accuracy across three datasets: MNIST, Fashion MNIST, and KMNIST. The methods include QAE classification, NBMF, FCNN, and SNN. Sparsity refers to the proportion of parameters within a model that are effectively inactive in learning}\label{tab:other_methods}%
\centering
\resizebox{\textwidth}{!}{%
\begin{tabular}{@{}ccccccc@{}}
\toprule
Method & Sparsity & Parameters & MNIST & Fashion MNIST & KMNIST & Average \\
\midrule
QAE classification & 0\% & 168 & 90.4\% & 67.2\% & 65.4\% & 74.3\% \\
NBMF & 74\% & 2198 & 91.0\% & 84.2\% & 75.4\% & 83.5\% \\
NBMF & 97\% & 247 & 23.0\% & 39.4\% & 33.0\% & 31.8\% \\
FCNN & 0\% & 8356 & 95.6\% & 89.4\% & 79.6\% & 88.2\% \\
SNN & 74\% & 2192 & 95.6\% & 89.2\% & 77.2\% & 87.3\% \\
SNN & 98\% & 174 & 57.4\% & 57.8\% & 43.4\% & 50.6\% \\
\botrule
\end{tabular}%
}
\end{table*}
FCNN achieved the highest average test accuracy, consistent with the well-established effectiveness of the fully-connected network in image classification. SNN model, in which the FCNN was made 74\% sparse, retained a sufficient number of trainable parameters. Similar to NBMF, achieved a higher average classification accuracy than QAE classification.
However, despite having equal feature dimensionality of the latent space, QAE classification significantly reduces the number of trainable parameters compared to both classical methods. The following percentages represent how much the number of parameters was reduced by QAE classification compared to other conventional models:
\begin{itemize}
\item 98.0\% compared to FCNN with 0\% sparsity
\item 92.4\% compared to NBMF with 74\% sparsity
\end{itemize}
Despite this significant parameter reduction, QAE classification maintains 9.2 and 13.9\% average accuracy reductions compared to NBMF and FCNN, respectively.
Conversely, when the number of parameters is adjusted to approximate that of QAE classification, the classification accuracy of NBMF and FCNN significantly decreases, further widening the performance gap with QAE classification.

The performance of QAE classification is compared with the results of QCNN under identical training and testing data conditions. The network configuration for QCNN is described in Section \ref{sec:QCNN}. Additionally, the results are compared with those of a classical CNN, where convolutional operations are performed using classical computation. For both QCNN and CNN, the results at convergence after 1000 training epochs are used. To ensure a fair comparison, the QAE-based classification results at 1000 epochs are also reported. Table~\ref{tab:QCNN} presents the test accuracy of each method on the same dataset.
\begin{table*}[t]
\caption{Comparison of different classification methods with respect to the number of parameters and test accuracy across MNIST, Fashion MNIST, and KMNIST datasets. The methods include QAE classification, QCNN, and CNN}
\label{tab:QCNN}
\centering
\resizebox{\textwidth}{!}{%
\begin{tabular}{@{}cccccc@{}}
\toprule
Method & Parameters & MNIST & Fashion MNIST & KMNIST & Average \\
\midrule
QAE classification & 168 & 80.6\% & 64.4\% & 53.2\% & 66.1\% \\
QCNN & 8372 & 94.8\% & 90.0\% & 82.2\% & 89.0\% \\
CNN & 8376 & 95.0\% & 90.2\% & 80.6\% & 88.6\% \\
\botrule
\end{tabular}%
}
\end{table*}
The test accuracy of QCNN and CNN exceeds that of QAE classification. Both methods achieve superior accuracy on the Fashion MNIST and KMNIST datasets, where QAE classification, as well as NBMF and FCNN in Table~\ref{tab:other_methods}, struggled to achieve high performance. However, despite the introduction of quantum computation in QCNN, its performance shows little to no significant improvement over the classical CNN. This is notable given that CNN already achieves high accuracy, raising questions about the actual benefits of integrating quantum computation in the QCNN architecture.

\section{Discussion}
To examine the relationship between the ansatz structure and classification performance, we refer to the results presented in Table~\ref{tab:ansatz_accuracy}. The classification accuracy varied depending on the dataset. Among the datasets, MNIST achieved the highest classification accuracy. This dataset contains images of handwritten digits. Despite variations in handwriting styles and stroke thickness, the digits generally exhibit continuous, single-stroke structures. The differences among data samples within the same class are minimal, making this dataset relatively less challenging for image recognition compared to others.
In contrast, KMNIST, while also a handwritten dataset, consists of images of traditional Japanese characters. The complexity of the character structures, combined with significant variations in individual writing styles, makes this dataset more challenging for classification than MNIST. Fashion MNIST, on the other hand, features images of fashion items centrally placed within the image. A distinctive characteristic of this dataset is the concentration of pixel values in the central region, differing significantly from the other datasets.

Variations in classification accuracy are observed not only due to differences in the datasets but also depending on the structure of the ansatz within the QAE classification circuit. These differences were reflected in the average accuracy values shown in Table~\ref{tab:ansatz_accuracy}. Although the number of trainable parameters was nearly identical across different ansatz configurations, their test accuracies differed. This suggests that the feature extraction enabled by the quantum gate configurations directly impacts learning performance.
Circuits 1 through 3 consistently achieved high classification accuracy across all datasets and are characterized by a simple structure that entangles qubits using only CNOT gates. Circuits 4 and 5 also achieved relatively high accuracy, utilizing CR gates to generate entanglement. As shown in Table~\ref{tab:gate_num}, Circuits 6 and 7, which employ CRX or CRZ gates and do not include any CNOT gates, have fewer two-qubit gates overall. These ansatz structures exhibited lower classification performance, suggesting the importance of entangling information between qubits.
Nevertheless, the results also indicate that simply using CNOT gates does not guarantee high accuracy. Circuit 5, which applies CNOT gates between all possible pairs of qubits, has significantly more CNOT gates than other ansatzes, as shown in Table~\ref{tab:gate_num}. Despite this, it exhibited substantially lower classification accuracy compared to the others. This suggests that simply increasing circuit depth by adding CNOT gates is insufficient for improving accuracy.
The superior performance of Circuit 3 over Circuit 1 further suggests that the arrangement of quantum gates is also a critical factor. While Circuit 1 arranges CNOT gates sequentially from lower to higher qubits, Circuit 3 first has the lowest qubit control the highest qubit, followed by a downward sequence of CNOT gates. Although these two circuits share an identical number and type of quantum gates, their classification accuracies differed, indicating that the placement of gates must be carefully considered.
In addition, the choice of rotation axes within the ansatz is also considered to influence classification performance. Circuit 7, which employs CRX gates, showed low accuracy. In this circuit, RX gates are fixed as single-qubit rotations, and the use of CRX gates results in state transformations constrained to the X-axis, limiting the expressive capability of the circuit. Similarly, although Circuit 5 utilizes CNOT gates, its rotation gates at both ends are also RX, effectively restricting the rotational operations to a single axis.
Circuits 6 and 7, which include CRZ gates, also exhibited low classification accuracy. In QAE classification, quantum states $\ket{0}$ and $\ket{1}$ are assigned to the trash state to represent class information. Applying rotations along the Z-axis to these states may not contribute meaningfully to their transformation, which likely accounts for the reduced performance observed in these circuits.

Overall, the results of this experiment demonstrate that the structure of the ansatz has a significant impact on classification accuracy. In particular, Circuits 1 through 3, which employ simple entanglement structures using only CNOT gates, achieved high performance. These findings suggest that the choice of gate types, their arrangement, and the design of rotation axes are critical factors directly affecting model performance. On the other hand, a greater number of gates or increased structural complexity does not necessarily lead to higher accuracy. These results indicate that careful circuit design is essential for performance optimization, and that even simple configurations can achieve sufficient accuracy when properly constructed.

The PCA plots in Figure~\ref{fig:pca} indicate that Circuit 3 achieves better classification performance than the other ansatz structures. For the remaining misclassified data points, the features appear ambiguous and difficult to distinguish, suggesting that such instances would pose challenges not only for QAE classification but also for conventional machine-learning methods.
In addition, the comparison of logarithmic loss in Table~\ref{tab:log_loss} indicates the impact of the ansatz structure on accuracy. Despite using the same dataset and a nearly identical number of optimized parameters, modifying the combination of quantum gates within the ansatz enhances prediction performance. A lower logarithmic loss suggests higher confidence in the model's predictions, further highlighting the effectiveness of specific ansatz configurations.
The expressibility value in Table~\ref{tab:exprs} also supports the other results. Among the tested ansatz, Circuit 3 exhibited the lowest $D_{KL}$ value, suggesting it possesses the broadest capacity to represent diverse quantum states. This implies that a wider range of quantum states can be effectively represented using CNOT gates alone, without the need for fine-grained angle adjustments via continuous parameters.
However, in the case of Circuit 4, where RY gates were assigned as arbitrary quantum gates, $D_{KL}$ was higher than that of other circuits, despite achieving relatively high average accuracy. This suggests that lower $D_{KL}$ does not necessarily lead to improved classification performance. Instances have been observed where classification accuracy plateaus even when $D_{KL}$ is low, indicating that there is not always a clear causal relationship between the two \citep{Hubregtsen_2021}. Expressibility should not be used as the sole indicator when evaluating quantum circuits, but rather considered as a supplementary metric alongside other experimental results.


Extending the QAE classification model from the four-class task (where it achieved 90.4\% test accuracy) to an eight-class task resulted in the test accuracy significantly dropping to 49.4\%. Initial attempts to improve performance mirrored common strategies in traditional machine learning: increasing the number of training samples, optimized parameters (by increasing ansatz repetitions), and the number of qubits in the latent space. While these modifications led to a modest improvement in test accuracy, they were insufficient to recover the performance levels observed in the four-class scenario. This suggests that, similar to classical machine learning, expanding the diversity of the feature space is beneficial but not a complete solution. 
It points to the need for more sophisticated improvements, such as refining the data encoding method or fundamentally restructuring the ansatz. Analysis of the confusion matrix (Figure~\ref{fig:prediction_heatmap}) reveals that certain digit classes exhibit significantly lower classification accuracy, suggesting a weakness in recognizing specific numerical shapes. This issue likely arises because the QAE classification architecture, as implemented in this experiment, fails to adequately capture the relevant features of these digit shapes.

While maintaining the learning mechanism for classification using QAE, two strategies are considered to improve multiclass classification accuracy.
The first strategy is to change the encoding method for image data. In this experiment, amplitude encoding was used. Although amplitude encoding can compress $O(2^n)$ classical data into $O(n)$ quantum states, it inevitably leads to the loss or inability to represent certain information. Encoding methods such as basis encoding or angle encoding, which represent one pixel of image data with one qubit, can express image information more accurately \citep{Rath_2024}. In addition, NEQR stores color information directly into the basis states, providing more reliable encoding compared to FRQI, which assigns information as amplitudes \citep{Zhang_2013, Phuc_Q_2011}. However, when using these methods, the number of qubits required for encoding becomes impractically large for current quantum computers. For example, to encode the $16 \times 16$ images used in this experiment with basis encoding or angle encoding, 256 qubits would be necessary. Considering that the Heron r2 processor, which has the largest number of qubits publicly available from IBM as of 2025, is equipped with 156 qubits, it is clear that this problem size remains difficult to calculate \citep{AbuGhanem_2025}.
Another potential avenue for improving digit shape recognition involves leveraging classical preprocessing techniques before QAE classification. For instance, PCA could be used to transform the raw image data into principal components, capturing the dominant features. Encoding these PCA-extracted features, rather than the raw pixel values, into quantum states might allow the QAE to begin with a more structured and informative representation of the digit shapes, potentially boosting classification accuracy. However, this approach introduces a classical preprocessing step, compromising the end-to-end quantum nature of the QAE classification and making it a less desirable solution from the perspective of exploring purely quantum methods. 

The second strategy is to modify or enhance the qubit interactions within the ansatz—perhaps by exploring different connectivity patterns or incorporating more complex gate sequences—could improve feature extraction for these challenging classes. The relative simplicity of the ansatz structure in Circuit 3, while beneficial for expressibility, might limit its ability to capture the intricate features of certain digit shapes. However, excessively increasing the number of parameters or introducing deeply hierarchical ansatz structures could negatively impact learning efficiency. Therefore, a careful balance between enhancing representational power and maintaining computational tractability is crucial when considering such modifications.
A possible method to improve the ansatz design is to apply Quantum Architecture Search (QAS) \citep{Wenjie_2023, Zhimin_2025}. QAS is a technique that searches for the optimal combination of quantum gates instead of fixing the gate structure of the ansatz in advance. An objective function is defined, and the search aims to find a combination of quantum gates that maximizes the fulfillment of this objective. For example, in QAE classification, by setting an objective function derived from classification accuracy, it is possible to search for an ansatz structure effective for high-accuracy classification. One concern with this method is that both the selection of gates and the optimization of ansatz parameters must be performed, leading to a higher computational cost compared to cases where the ansatz structure is predetermined. The use of QAS will be considered by taking into account the trade-off between its contribution to improving accuracy and the resulting computational cost.

As confirmed in Table~\ref{tab:other_methods}, comparing QAE classification with classical machine-learning methods highlights a key advantage: QAE classification achieves comparable test accuracy while substantially reducing the number of optimized parameters. When utilizing amplitude encoding, the initial 256-dimensional classical image data ($16 \times 16$ pixels) is compressed into an eight-qubit quantum state (and further down to a five-qubit latent space). In contrast, the conventional methods employed 32-dimensional feature vectors. Despite this significant reduction in both parameters and the dimensionality of the feature representation, QAE classification maintains high accuracy. This strongly suggests that QCL, even with a relatively simple ansatz, can efficiently extract essential features for classification. By transferring the nonlocal observable to a single-qubit observable using entangling gate such as the CNOT gate, QCL efficiently extracts high-order polynomial terms into a measurable form \citep{Mitarai_2018}. The tensor product structure of quantum systems readily calculates the product of input states, which may enable learning of complex functions with few parameters.
While NBMF, another comparative method, utilizes annealing-based (potentially quantum-accelerated) computation, the results of this study demonstrate the effectiveness of gate-based QCL, indicating the broader potential of quantum algorithms in the machine-learning pipeline. 

By implementing two strategies—modifying the encoding method and exploring different ansatz structures—QAE classification is expected to retain substantial potential for improving classification accuracy. These enhancements will become feasible as the number of qubits available on quantum computers increases and as quantum computations become executable within practical timeframes. Moreover, based on the comparative results presented in Table~\ref{tab:other_methods}, it is anticipated that even with these improvements, the number of parameters to be optimized can be significantly reduced compared to classical machine-learning methods. In the future, where gate-based quantum computers are more advanced, QAE classification may offer distinct advantages over classical approaches for multiclass classification tasks.

However, it is important to acknowledge that the classical FCNN achieves superior test accuracy. From a practical standpoint, FCNNs operating on classical hardware are far more accessible and resource-efficient than QAE classification, which relies on quantum simulators or, ideally, future fault-tolerant quantum computers. Additionally, training convergence is significantly faster with classical methods. Nevertheless, the current computational disadvantage of QCL may diminish with future advancements in gate-based quantum-computing hardware. A year-by-year roadmap projects a steady increase in the number of qubits installed in gate-based quantum computers, envisioning that by around 2033, such systems will be capable of executing circuits with one billion gates using up to 2000 qubits \citep{AbuGhanem_2025, IBM_roadmap}. If such large-scale hardware is achieved in the future, it will become feasible to measure large quantum circuits within practical computation times. For the proposed method, this development may not only accelerate training but also enhance the encoding performance and resolution of image data. As a result, it may enable the provision of more accurate and practical QAE classification models. In addition, the concept of ``quantum-centric supercomputing" has also been proposed \citep{AbuGhanem_2025, Alexeev_2024}. By integrating with high-performance computing systems, it is expected to accelerate the processing speed of quantum computing. Although applications in materials science have been demonstrated, it is also conceivable that this approach may contribute to the acceleration of training and the enhancement of data resolution in the proposed method \citep{Alexeev_2024}.

As indicated by the results in Table~\ref{tab:QCNN}, both QCNN and CNN achieved high classification accuracy not only on the MNIST dataset but also on Fashion MNIST and KMNIST. Compared to other machine learning methods, this demonstrates that convolutional processing effectively captures various data patterns. However, there is little to no performance difference between QCNN and classical CNN. While this could be interpreted as QCNN achieving performance comparable to its classical counterpart, the reality is that, as described in Section \ref{sec:QCNN}, quantum convolutional layers are applied only to a portion of the network. As a result, the benefits of quantum machine learning are minimally realized. Nevertheless, this network design represents the most feasible and practical approach for NISQ era quantum computers. Designing a fully quantum QCNN, where the entire network is constructed using quantum circuits and classification is performed via quantum measurement, would require a large number of qubits and quantum gates \citep{Cong_2019, Hur_2022}. In contrast, QAE-based classification offers a significant advantage in that it can perform both feature learning and multi-class classification using a small-scale parameterized quantum circuit, as evidenced by the number of parameters in Table~\ref{tab:QCNN}.

\section{Conclusion}
This study introduced a novel approach to image classification using QAEs, extending their applicability beyond conventional data compression and reconstruction tasks. The results demonstrated that QAE-based classification achieved high accuracy in a four-class classification task, comparable to conventional machine-learning methods, while substantially reducing the number of parameters requiring optimization. Furthermore, the importance of quantum-gate structure in the ansatz was highlighted—particularly, simple CNOT-based real-amplitude circuits outperformed more complex configurations, indicating that circuit design has a critical impact on learning performance.
Compared with classical models, QAE classification retained competitive accuracy with dramatically fewer parameters. While FCNNs and NBMF performed well, QAE classification offered better parameter efficiency. Additionally, unlike QCNNs, which rely heavily on classical components, the proposed method emphasizes fully quantum processing, making it suitable for end-to-end quantum learning.

However, the model’s scalability to multiclass classification remains a challenge. As the number of classes increased, accuracy declined significantly, and even with additional training data or deeper circuits, recovery was limited. This suggests that more expressive feature encoding or advanced ansatz structures may be needed. Future improvements could involve alternative encoding methods such as angle encoding, or adaptive circuit design through Quantum Architecture Search (QAS), although these come with greater resource costs.

Although current hardware limitations and simulation costs restrict practical implementation, ongoing advances in gate-based quantum computing may unlock the potential of QAE classification for real-world applications. Overall, this study presents an efficient and compact quantum model for image classification, providing a foundation for further research toward scalable and hardware-aware quantum learning systems.

\backmatter

\bmhead{Acknowledgements}
The authors thank Editage (www.editage.jp) for English language editing.

\bmhead{Funding}
This study was partially supported by JSPS KAKENHI Grant Number JP23H04499, JP25H01522, and the Ochanomizu University Graduate Student Research Grant for FY2024.

\bmhead{Data Availability}
The processed datasets generated during the current study are available on GitHub at https://github.com/asaocha/QAE-classifier.git. Raw data and additional materials are available from the corresponding author upon reasonable request.

\bmhead{Author contributions}
H.A. conceived and conducted the experiments and analyzed the results.
K.K. supervised the project.
All the authors wrote and reviewed the manuscript.

\section*{Declarations}

\bmhead{Conflict of interest}
The authors declare no competing interests.

\bibliography{sn-bibliography}

\end{document}